\documentclass[superscriptaddress,showkeys,nofootinbib,aps,prd,11pt]{revtex4-1}
\usepackage{amsmath}
\usepackage{amssymb}
\usepackage{graphicx}
\usepackage{color}
\usepackage{float}
\usepackage{hyperref}

\newcommand{\braket}[1]{\left\langle#1\right\rangle}

\begin{document}
\title{\mathversion{bold}Unified SU(4) theory for the $R_{D^{(*)}}$ and $R_{K^{(*)}}$ anomalies}

\author{Shyam Balaji} 
\email{shyam.balaji@sydney.edu.au}
\affiliation{ARC Centre of Excellence for Particle Physics at the Terascale, School of Physics, The University of Sydney, NSW 2006, Australia}

\author{Michael A.~Schmidt} 
\email{m.schmidt@unsw.edu.au}
\affiliation{School of Physics, The University of New South Wales, Sydney, NSW 2052, Australia}

\begin{abstract}
We propose a chiral Pati-Salam theory based on the gauge group $SU(4)_C\times SU(2)_L\times SU(2)_R$. The left-handed quarks and leptons are unified into a fundamental representation of $SU(4)_C$, while right-handed quarks and leptons have a separate treatment. The deviations measured in the rare semileptonic decays $B\to D^{(*)} \tau \bar \nu$ are explained by a scalar leptoquark which couples to right-handed fields and is contained in the $SU(4)_C\times SU(2)_R$-breaking scalar multiplet. The measured deviation of lepton flavor universality in the rare decays $\bar B \to \bar K^{(*)} \ell^+ \ell^-$, $\ell = \mu, e$ is explained via the $SU(4)_C$ leptoquark gauge boson. The model predicts a new sub-GeV scale sterile neutrino which participates in the anomaly and can be searched for in upcoming neutrino experiments. The theory satisfies the current most sensitive experimental constraints and its allowable parameter regions will be probed as more precise measurements from the LHCb and Belle II experiments become available.
\end{abstract}


\maketitle


\section{Introduction}

The standard model (SM) of particle physics with the inclusion of neutrino masses describes nature with unprecedented precision and has so far withstood all experimental tests. However, recently several hints for a violation of lepton flavor universality (LFU) 
in recent measurements of semileptonic $B$ meson decays~\cite{Aaij:2014ora,Aaij:2017vbb,Aaij:2013qta,Aaij:2015oid,Wehle:2016yoi,Aaij:2014pli,Aaij:2015esa,Aaij:2019wad} have emerged. 
The theoretically cleanest probes are the LFU ratios
\begin{equation}\label{eq:RKRD}
    R_{K^{(*)}} = \frac{\Gamma(\bar B\to \bar K^{(*)} \mu^+\mu^-)}{\Gamma(\bar B\to \bar K^{(*)}e^+e^-)}
\qquad\text{and}\qquad
    R_{D^{(*)}} = \frac{\Gamma(B\to D^{(*)} \tau \bar \nu)}{\Gamma( B\to D^{(*)} \ell \bar \nu)}\;,
\end{equation}
where $\ell$ is a light lepton $\ell=e,\mu$, because hadronic uncertainties cancel out in the LFU ratios~\cite{Hiller:2003js}. Their current experimental measurements and SM predictions are summarized in Table~\ref{tab:RKRD}. While the LFU ratios $R_{K^{(*)}}$ point to a smaller decay rate with final state muons compared to electrons in the neutral current process $b\to s \ell^+\ell^-$, the LFU ratios $R_{D^{(*)}}$ indicate an enhanced rate for the charged current process $b\to c\tau\bar\nu$ compared to light charged leptons in the final state.
The significance of the anomalies in semileptonic $B$ meson decays is 
at the level of $2.5\sigma$ for both $R_K$ and $R_{K^*}$ ratios, while the significance for the combined measurement of the LFU ratios $R_D$ and $R_{D^*}$ exceeds $3\sigma$. 
\begin{table}[bh!]\centering
	\begin{ruledtabular}
		\begin{tabular}{cccc}
		&  Observed & SM & $q^2$ range\\ \hline
		$R_{K}$ & $0.846 ^{+0.060} _{-0.054}{}^{+0.016} _{-0.014}$ \cite{Aaij:2019wad}	&	$1.0003\pm 0.0001$ \cite{Bobeth:2007dw} & $1\, \mathrm{GeV}^2 < q^2<6\, \mathrm{GeV}^2$  \\
		$R_{K^*}$ 
		&$0.685 ^{+0.113} _{-0.069}\pm 0.047$ \cite{Aaij:2017vbb}
		& $1.00 \pm 0.01$ \cite{Bordone:2016gaq}
		& $1.1\, \mathrm{GeV}^2 < q^2 < 6\, \mathrm{GeV}^2$ \\
		$R_{D}$ & $0.340 \pm {0.027} \pm 0.013$ \cite{HFLAV18}	&	$0.299\pm 0.011$ \cite{Lattice:2015rga} & Full  \\
		$R_{D^{*}}$ & $0.295 \pm 0.012 \pm 0.008$ \cite{HFLAV18}	&	$0.252\pm 0.003$ \cite{Fajfer:2012vx} & Full  \\		
\end{tabular}
\end{ruledtabular}
\caption{Experimental results and standard model theory predictions for the LFU ratios $R_{K^{(*)}}$ and $R_{D^{(*)}}$. Statistical uncertainties are listed first and systematic uncertainties second. In the case of the LFU ratios $R_{K^{(*)}}$, the data are binned in the invariant mass $q^2$ of the final state lepton pair, in order to avoid the $J/\psi$ and other resonances. The relevant $q^2$ range is indicated in the last column.}
\label{tab:RKRD}
\end{table}

The experimental anomalies in $R_D$ and $R_{D^*}$ are supported by a similar deviation in the LFU ratio $R_{J/\psi}=\Gamma(B_c^+\to J/\psi \tau^+\nu)/\Gamma(B_c^+\to J/\psi \mu^+\nu)$ which analogously points to a larger branching fraction to tau leptons compared to muons, although still being consistent with the SM at the $2\sigma$ level due to large experimental uncertainties~\cite{Aaij:2017tyk}. Also, there are 
deviations in the angular observable $P_5^\prime$~\cite{Descotes-Genon:2013wba,Guadagnoli:2017jcl} and more generally data from several measurements of $b\to s\mu^+\mu^-$~\cite{Descotes-Genon:2015uva} that suggest a suppression of the decays $b\to s\mu^+\mu^-$ compared to the SM expectation, while being consistent with the experimentally observed value of the LFU ratios $R_{K^{(*)}}$. However, as these other channels currently have fewer clean signals due to large hadronic uncertainties in absolute branching ratio measurements and due to the difficulty of estimating a signal for $P_5^\prime$~\cite{Guadagnoli:2017jcl} along with other experimental uncertainties, we instead focus on the LFU ratios introduced in Eq.~\eqref{eq:RKRD} in the following discussion.

The possibility that some or even all of these deviations might be a harbinger of new physics has been entertained in the literature.
In particular, several $SU(4)$ models~\cite{Assad:2017iib,DiLuzio:2017vat,Calibbi:2017qbu,Bordone:2017bld,Barbieri:2017tuq,Blanke:2018sro,Greljo:2018tuh,Bordone:2018nbg,Faber:2018qon,Heeck:2018ntp,Balaji:2018zna,Fornal:2018dqn} have been proposed.
Most of these models simultaneously explain the $B$ physics anomalies via a massive vector leptoquark $W^\prime\sim (3,1,4/3)$\footnote{$W^\prime$ is the $U_1$ vector leptoquark in the nomenclature of Ref.~\cite{Buchmuller:1986zs}.}, which is predicted by the breaking of $SU(4)_C \to SU(3)_C$. In particular chiral $SU(4)$ models~\cite{Buttazzo:2017ixm,Balaji:2018zna,Fornal:2018dqn} are phenomenologically motivated, because they avoid constraints from lepton-flavor-violating pseudoscalar meson decays like $K_L\to e^\pm \mu ^\mp$, which place stringent constraints on the $SU(4)$-breaking scale~\cite{Hung:1981pd,Valencia:1994cj,Smirnov:2007hv,Kuznetsov:2012ai,Carpentier:2010ue,Smirnov:2018ske}. The authors of Ref.~\cite{Hati:2019ufv} find that minimal models with a single vector leptoquark and a unitary quark-lepton mixing matrix are generally disfavored due to strong constraints from charged lepton-flavor-violating processes. 

We pursue a different approach and build on our previously suggested Pati-Salam inspired chiral $SU(4)$ gauge model~\cite{Balaji:2018zna}, where the $b\to s$ anomaly is explained via the vector leptoquark $W^\prime$ with purely left-handed couplings. The explanation of $R_{K^{(*)}}$ is predictive and depends on only two parameters, the mass of the vector leptoquark and a CKM-type mixing angle between left-handed down-type quarks and charged leptons. One interesting feature of the model is the simultaneous modification of both the decay to muons, $b\to s \mu^+\mu^-$, and electrons\footnote{Modifications to electrons have been suggested in Ref.~\cite{Hiller:2014ula} and also realized in the simultaneous explanation of both anomalies using the $R_2$ leptoquark~\cite{Popov:2019tyc}.}, $b\to s e^+ e^-$, in opposite directions by an equal amount.
Here, we consider a simple extension of the model with a larger gauge group $SU(4)_C\times SU(2)_L\times SU(2)_R$ which further unifies the right-handed matter fields. This allows explanation of the $R_{D^{(*)}}$ anomalies with a new scalar leptoquark $\tilde \chi\sim(3,1,-2/3)$\footnote{$\tilde\chi$ corresponds to the conjugate of the $S_1$ leptoquark in the nomenclature of Ref.~\cite{Buchmuller:1986zs}.\label{foot:S1}}, which is part of the scalar breaking of the Pati-Salam gauge group to the SM gauge group.
The $\tilde \chi$ leptoquark is well known as an explanation for $R_{D^{(*)}}$~\cite{Sakaki:2013bfa,Freytsis:2015qca} and other hints of new physics (see e.g.~Refs.~\cite{Bauer:2015knc,Popov:2016fzr,Cai:2017wry,Azatov:2018kzb,Bigaran:2019bqv}).
Here, the $\tilde\chi$ leptoquark features purely right-handed couplings and thus mediates $b_R\to c_R\tau_R \nu$ with right-handed charged fermions, where $\nu$ is a new light sterile neutrino. The sterile neutrino can be searched for and provides a smoking-gun signature of the explanation of the observed measurement of $R_{D^{(*)}}$. 

The paper is structured as follows. In Sec.~\ref{sec:model} we introduce the model and discuss the scalar potential and fermion masses. New contributions to the $B$ physics anomalies are discussed in Sec.~\ref{sec:anomalies} and relevant constraints in Sec.~\ref{sec:constraints}. In Sec.~\ref{sec:results} we present our results before concluding in Sec.~\ref{sec:conclusions}. The decomposition of the particle content in terms of SM multiplets is shown in the appendix.


\section{Model}
\label{sec:model}
We propose a model based on the gauge group $SU(4)_C\times SU(2)_L\times SU(2)_{R}$ and assign the particle content such that the $SU(4)$ leptoquark gauge boson couples to the quarks and leptons in a chiral fashion. This naturally avoids strong constraints from charged lepton-flavor-violating leptonic neutral meson decays such as $K_L\to e^\pm \mu^\mp$ and $B\to e^\pm \mu^\mp$. The particle content of the model is listed in Table~\ref{tab:particles}.
\begin{table}[bht!]
        \centering
	\begin{ruledtabular}
        \begin{tabular}{ccc|cc}
		Fermion & $(SU(4)_C, SU(2)_L, SU(2)_{R})$ & Generations
			& Scalar & $(SU(4)_C,SU(2)_L,SU(2)_{R}$)
                \\\hline
                ${\bf Q}_L$ & $(4,2,1)$ & 3 & 
                $\phi$ & $(1,2,2)$\\
                ${\bf Q}_R$ & $(4,1,2)$ & 3 &
                $\chi$ & $(4,1,2)$\\
                ${\bf f}_R$ & $(1,1,3)$ & 3& 
                $\Delta$ & $(4,2,3)$\\
                $S_L$&$(1,1,1)$ &1 
        \end{tabular}
\end{ruledtabular}
\caption{Particle content}
        \label{tab:particles}
\end{table}
Apart from the usual matter fields ${\bf Q}_{R,L}$ in the fundamental representation of $SU(4)_C$, there are three generations of right-handed triplet fermions ${\bf f}_R$ and a left-handed total singlet fermion $S_L$. The scalar sector consists of a bidoublet $\phi$ and two fields in the fundamental representation of $SU(4)_C$, $\chi$ and $\Delta$.

The $SU(4)_C\times SU(2)_R$ symmetry is broken by the vacuum expectation value (VEV) of the scalar $\chi$ at a high scale, $\langle \chi_{41}\rangle \equiv w \gtrsim 20$ TeV, where the first (second) index refers to the fundamental representation of $SU(4)_C$ ($SU(2)_R$). Electroweak symmetry is broken by
the scalar $\phi$ with $v\equiv \sqrt{|v_{12}|^2 +|v_{21}|^2} \simeq  (2\sqrt{2} G_F)^{-1/2} \simeq174 \ {\rm GeV}$ where  $v_{12}\equiv \langle \phi_{12}\rangle$ and $v_{21}\equiv \langle \phi_{21} \rangle$ refer to the VEVs in the $(I_{3L},I_{3R})=(\frac{1}{2},-\frac{1}{2})$ and ($I_{3L},I_{3R})=(-\frac{1}{2}, \frac{1}{2})$ components. 
The combination of the VEVs of $\chi$ and $\phi$ induces small VEVs for $\Delta$, $\langle\Delta_{41(12)}\rangle = u_1$ and $\langle \Delta_{42(11)}\rangle =u_2$. The first index refers to the fundamental representation of $SU(4)_C$, the second refers to the fundamental representation of $SU(2)_L$ and the last two in round brackets are two indices in the fundamental representation of $SU(2)_R$ which are symmetrized as indicated by the round brackets, $T_{(ab)}=\tfrac12(T_{ab}  + T_{ba})$. Thus the following symmetry-breaking pattern emerges $|u_1|^2+ |u_2|^2\ll |v_{12}|^2+|v_{21}|^2 \ll |w|^2$
\begin{equation}
\begin{gathered}
    SU(4)_C \times SU(2)_L \times SU(2)_{R}  \\
   \downarrow \braket{\chi}\\
   SU(3) \times SU(2)_L \times U(1)_{Y} \\
   \downarrow \braket{\phi}, \braket{\Delta}\\
   SU(3) \times  U(1)_Q \\
\end{gathered}
\end{equation}
Here weak hypercharge $Y$ and electric charge $Q$ are related to the generators in $SU(4)_C\times SU(2)_L\times SU(2)_R$ by $Y=T+2 I_{3R}$ and $Q=\frac{T}{2}+I_{3L}+I_{3R} = I_{3L}+Y/2$, respectively.
If we use the gauge symmetry to rotate the VEV of $\chi$ to the fourth component, then $T$ is the diagonal traceless $SU(4)$ generator with elements $(\frac{1}{3},\frac{1}{3},\frac{1}{3},-1)$.

\subsection{Yukawa sector} \label{sec:YukawaSector}
Given the particle content in Table~\ref{tab:particles} the full Yukawa Lagrangian is given by 
\begin{align}\label{eq:Yukawa}
	\mathcal{L} =&
        Y_1 \bar {\bf Q}_L^{ia} \phi_{a\beta} ({\bf Q}_R)_{i\gamma}\varepsilon^{\beta \gamma}
        + Y_2 \bar {\bf Q}_L^{ia}  \tilde \phi_{a \beta} ({\bf Q}_R)_{i\gamma}\varepsilon^{\beta \gamma}
        - Y_3 \bar {\bf Q}_R^{i\alpha} \chi_{i\alpha} {S}_L
	+ Y_4 \bar {\bf Q}_R^{i\alpha} \chi_{i\beta} ({\bf f}^c_R)_{(\alpha\gamma)}\varepsilon^{\beta \gamma} 
\\\nonumber &
        + Y_5 \bar {\bf Q}_L^{ia} \Delta_{ia(\alpha\beta)} ({\bf f}_R)_{(\gamma\delta)}\varepsilon^{\alpha\gamma}\varepsilon^{\beta\delta}
        + \frac12 m (\bar {\bf f}_R^c )^{(\alpha\beta)}({\bf f}_R)_{(\alpha\beta)}
	- \frac12 m_S S_L^T \hat{\bf C} S_L
        +h.c.\;,
\end{align}
where flavor indices are suppressed, but indices for the gauge groups are explicitly shown.\footnote{Lower indices refer to the fundamental representation and upper indices refer to the antifundamental representation. Fields $\psi$ with lower indices transform as $\psi_i\to (U\psi)_i \equiv U_i^j \psi_j$.} The Yukawa couplings are matrices in flavor space; rows (columns) are labeled by the first (second) fermion in the fermion bilinear. Indices in fundamental representation of $SU(4)_C$ are labeled by roman letters $i,j,\dots$, indices in the fundamental representation of $SU(2)_L$ are labeled by greek letters $\alpha,\beta,\dots$ and indices in the fundamental representation of $SU(2)_R$ are labeled by roman letters $a,b,\dots$.
In the above expression we used the charge-conjugate fields $\tilde \phi_{\alpha a} = \epsilon_{\alpha \beta} \epsilon_{ab} \phi^{*\beta b}$ and $(f_R^c)_{\alpha\beta} = \epsilon_{\alpha\alpha^\prime}\epsilon_{\beta\beta^\prime} \hat{\bf C} \gamma^0 f_R^{*\alpha^\prime \beta^\prime}$ with $\hat{\bf C}=i\gamma^2\gamma^0$.
From the Yukawa Lagrangian \eqref{eq:Yukawa} we obtain the Lagrangian of the quark masses
\begin{align}
    \mathcal{L} = 
    - \bar {\bf u}_L m_u {\bf u}_R
    - \bar {\bf d}_L m_d {\bf d}_R + h.c.
\end{align}
with the quark mass matrices 
\begin{align}
    m_u & = Y_1 v_{12} +Y_2 v_{21}^* &
    m_d & = - Y_1 v_{21} - Y_2 v_{12}^* \;.
\end{align}
The charged and neutral lepton mass matrices can be written in the basis 
\begin{align}
	\mathcal{L}& = 
	-\frac12 \mathcal{N}^T\hat{\bf C}M_{\nu,N} \mathcal{N}
	- \left(\bar{\mathcal{E}}_L M_{e,E} \mathcal{E}_R +h.c.\right)
	&
	\mathcal{E}_L &\equiv \begin{pmatrix} \mathbf{e}_L\\ \mathbf{E}_L\end{pmatrix} &
	\mathcal{E}_R &\equiv \begin{pmatrix} \mathbf{e}_R\\\mathbf{E}_R\end{pmatrix} &
	\mathcal{N} &\equiv \begin{pmatrix} \mathbf{\nu}_L\\\mathbf{\nu}_R^c\\\mathbf{N}_R^c\\S_L\end{pmatrix}
\end{align}
with the mass matrices 
\begin{align}
    M_{e,E} &= \begin{pmatrix}
     -Y_5 u_2 &  m_d\\
     -m & - Y_4^\dagger w^*
     \end{pmatrix} &
    M_{\nu,N} &= \begin{pmatrix}
    0 & m_u^* & \sqrt{2} Y_5^* u_1^* & 0\\
    . & 0 & -\frac{Y_4 w}{\sqrt{2}} &  Y_3 w\\
    . & . & -m^* & 0 \\
    . & . & . &m_S
    \end{pmatrix}
\end{align}
A viable mass spectrum for the charged leptons is obtained for $m_d, m \ll Y_4 w$. More precisely, we take the eigenvalues of $m$ to be less than $\simeq 1$ GeV and the eigenvalues of $Y_4 w$ to be larger than $\simeq1$ TeV. In this case the new charged fermions $E_{L,R}$ decouple and their masses are determined by $M_E\approx -Y_4^\dagger w^*$, while the light charged lepton masses are determined by $M_e \approx -Y_5 u_2$. The contribution from mixing with $E_{L,R}$ can be neglected because of the assumed relative sizes of $m$, $m_d$ and $Y_4 w$. In the basis of a diagonal $Y_5$ the SM charged lepton mass eigenstates are approximately given by the weak interaction eigenstates. We thus denote them by $e_{L,R}$. Neutrino mass eigenstates are labeled by $n_i$. Hence in this basis the leptonic mixing 
 matrix is determined by the neutrino mass matrix up to subpercent-level corrections from mixing with the heavy charged leptons.

The neutrino oscillation data and the existence of a fourth light sterile neutrino with $m_4\lesssim 1$GeV requires $u_1\ll u_2$ and $Y_4 w, Y_3 w \gg m, m_S$ to be satisfied. For the remainder of this work we focus exclusively on the limit $u_1\to0$ in order to recover the experimentally observed active neutrino mass spectrum and the leptonic mixing angles. In this limit, three pseudo-Dirac pairs obtain masses of order $Y_{3,4}w$ and decouple from four light neutrinos. A minimal phenomenologically viable texture for the neutrino mass matrix is given by 
\begin{align}\label{eq:minYuk}
    Y_3 & = \begin{pmatrix}
    0 \\ 0 \\ y_3
    \end{pmatrix} &
    Y_4& =\begin{pmatrix}
    Y_{ue} &0&0\\
    0& 0 & Y_{c\tau}\\
    0&Y_{t\mu} &0\\
    \end{pmatrix}\;.
\end{align}
The large off-diagonal entries $Y_{c\tau}$ and $Y_{t\mu}$ are required for the $b\to c$ anomalies. The entries of the Majorana mass matrices $m$ and $m_S$ have to be small $\lesssim 1$ GeV in order to kinematically allow $R_{D^{(*)}}$ from the relevant $b\to c\tau n_4$ process.


\subsection{Scalar potential}
In this model, the masses of the charged leptons arise from the VEV of the $\Delta$ scalar, while the masses of the quarks result from the VEVs of the bidoublet $\phi$. In such a situation, consistent Higgs phenomenology requires a decoupling limit where the LHC Higgs-like scalar is identified with the lightest neutral scalar in the model. The decoupling limit works analogously to the one shown in Refs.~\cite{Haber:1989xc,Balaji:2018zna} and thus, we do not repeat the whole discussion, but instead focus only on the pertinent differences in the following. 

In order to achieve the desired symmetry-breaking pattern, we first neglect the scalar $\Delta$ and focus on the scalars $\chi$ and $\phi$. In this case the possible invariants which enter the scalar  potential are
\begin{align}
    I_1&=\chi^{*i\alpha}\chi_{i\alpha} -w^2 &
    I_2&=\chi_{i\alpha} \chi_{j\beta}\chi_{k\gamma}\chi_{l\delta} \epsilon^{ijkl} (\epsilon^{\alpha\beta}\epsilon^{\gamma\delta} +\epsilon^{\alpha\gamma}\epsilon^{\delta\beta} + \epsilon^{\alpha\delta} \epsilon^{\beta\gamma}) +\mathrm{h.c.} \nonumber\\ 
    J_1&=\phi^{*a \beta}\phi_{a \beta} -(|v_{12}|^2+|v_{21}|^2) &
   J_2 &=\frac14 (\phi_{a\alpha} \phi_{b\beta} \epsilon^{ab}\epsilon^{\alpha\beta}+ \textrm{h.c.}) +\mathrm{Re}(v_{12}v_{21})\label{eq:invariants}\\\nonumber
    K_1 & = \left(\chi^{*i\alpha} \chi_{i\beta}-w^2\right) \left(\phi_{a\alpha}\phi^{*a\beta}-|v_{21}|^2\right) &
    J_3 &=\frac{1}{4i}(\phi_{a\alpha} \phi_{b\beta} \epsilon^{ab}\epsilon^{\alpha\beta}- \textrm{h.c.}) +\mathrm{Im}(v_{12}v_{21})
    \;,
\end{align}
where we have subtracted the VEVs from each invariant such that the invariants vanish in the vacuum. The VEV of $\chi$ can always be chosen to be real by using a suitable global $SU(4)_C\times SU(2)_R$ rotation. 
The terms $I_1$, $J_1$ and $K_1$ respect an accidental $U(1)_\chi\times U(1)_\phi$ symmetry. $U(1)_\chi$ is broken by $I_2$ and $U(1)_\phi$ is broken by $J_{2,3}$. The invariants $I_1$, $J_1$ and $K_1$ are non-negative, while the others may become negative and thus terms involving these have to be sufficiently small to ensure vacuum stability. As the discussion of the $B$ physics anomalies is mostly independent to the exact form of the scalar potential, we only comment on how to obtain the correct vacuum structure. The scalar potential in terms of invariants is given by
\begin{align}
V(\chi,\phi) 
=\lambda_1 I_1^2 + \lambda_2 I_2
+\frac12 \sum_{i=1}^3 \sum_{j=1}^i \lambda_{ij} J_i J_j
+\sum_{i=1}^3 \lambda^\prime_{i} I_1 J_i
+\lambda_4^\prime K_1
\;.
\end{align}
The coefficients $\lambda_i^\prime$ parametrize interactions between the $\chi$ and $\phi$ fields. 
Most of the scalar potential is invariant under a larger symmetry group $SU(4)_C\times SU(2)_L\times SU(2)_{R,\chi}\times SU(2)_{R,\phi}$ with two separate $SU(2)_R$ symmetries for each of the two scalars $\chi$ and $\phi$. It is only broken to the diagonal subgroup by the last term $\lambda_4^\prime K_1$. 
The couplings in the scalar potential can be chosen real due to the invariants in Eq.~\eqref{eq:invariants} being Hermitian. We also restrict ourselves to real VEVs. 
This potential allows the VEV hierarchy $w\gg v_{21}\gg v_{12}=0$ to emerge, which leads to the correct quark mass spectrum with $Y_2 = m_u/v_{21}^*$ and $Y_1 = -m_d/v_{21}$ as mentioned in the previous section. The scalar doublet in the bidoublet which does not obtain a VEV induces flavor changing neutral currents~\cite{Mohapatra:1983ae,Gilman:1983ce, Gilman:1983bh, Ecker:1983uh} which poses a lower bound on its mass scale of $\mathcal{O}(20)$ TeV~\cite{Bertolini:2014sua}.

There are many terms in the scalar potential which couple the scalar field $\Delta$ to other scalar fields. However most of them are not relevant for the induced VEVs of $\Delta$. The most important term is linear in $\Delta$ 
\begin{align}\label{eq:m123}
    V(\Delta,\phi,\chi) = m_{123}\, (\Delta^{*ia(\alpha\beta)}\, \phi_{a\alpha}\, \chi_{i\beta} + \mathrm{h.c.}) 
    = \sqrt{2} m_{123}\, v_{21}\, w\, h_3 +\dots \;,
\end{align}
where we have absorbed the phase of $m_{123}$ by rephasing $\Delta$ and defined the electrically neutral scalar $h_3\equiv\sqrt{2}\mathrm{Re}(\Delta_{42(11)})$. In order to calculate the induced VEV of the scalar $\Delta$ it is sufficient to consider terms quadratic in $h_3$, because the induced VEV is much smaller compared to all other scales. Thus we obtain
\begin{equation}
    u_2 \equiv \langle \mathrm{Re}(\Delta_{42(11)})\rangle = \frac{\langle h_3\rangle}{\sqrt{2}}= - \frac{m_{123}\,v_{21}\,w}{m_{h_3}^2}\;,
\end{equation}
where $m_{h_3}$ is the mass of $h_3$. In the limit $w^2\gg v_{21}^2$ the observed Higgs boson $h$ is a linear combination of $h_1\equiv \sqrt{2}\mathrm{Re}(\phi_{21})$ and $h_3$
\begin{equation}
    h = \cos\beta \, h_1  + \sin\beta \,h_3\;.
\end{equation}
The mixing arises from the term in Eq.~\eqref{eq:m123} and indirectly from terms quadratic in $\Delta$ and $\phi$, after $\Delta$ obtains a VEV $u_2$. Generally the mixing angle is given by $\sin\beta \sim m_{123} w/m_{h_3}^2 = u_2/v_{21}$ and thus the Higgs $h$ features SM-like couplings, as discussed in Refs.~\cite{Haber:1989xc,Balaji:2018zna}.


\section{New contributions to semileptonic $B$ decays}
\label{sec:anomalies}

In Ref.~\cite{Balaji:2018zna}, we showed that the experimentally observed
values of $R_K$ and $R_{K^*}$ can be explained via the exchange of the massive
leptoquark gauge boson $W^\prime$ in $SU(4)_C$. There has been a recent measurement of
$R_K$ by the LHCb experiment~\cite{Aaij:2019wad} (see Table~\ref{tab:RKRD}) and the LHCb experiment also
published a new stronger limit~\cite{Aaij:2019nmj} on the branching ratio of
the semileptonic charged lepton flavor violating decay $B\to K e^\pm \mu^\mp$: BR$(B^+\to K^+\mu^- e^+) <7.0\times 10^{-9}$ and BR$(B^+\to K^+ \mu^+ e^-)< 6.4 \times 10^{-9}$ at 90\% C.L. Hence we briefly summarize the relevant definitions in Sec.~\ref{sec:bsll} and update the analysis with the latest measurements in Sec.~\ref{sec:constraints}.

The aforementioned vector leptoquark $W^\prime$ cannot explain the measurement of $R_D$
and $R_{D^*}$ due to its chiral couplings. This model also features several scalar leptoquarks which also contribute to $R_{D^{(*)}}$: (i) The scalar
$\Delta$ contains two leptoquarks $\Delta_{i\alpha11}$ and $\Delta_{i\alpha(12)}$, denoted by $R_2$ and $\tilde R_2$ in the nomenclature of Ref.~\cite{Buchmuller:1986zs}. 
However these two leptoquarks have
chiral couplings and either couple to charged leptons or neutrinos, but not both simultaneously. Although their electric charge $2/3$ components mix and thus in general contribute to $R_{D^{(*)}}$, their contribution is suppressed due to the small mixing and thus cannot account for the observed deviation in $R_{D^{(*)}}$. 
(ii) The scalar $\chi$ also contains a leptoquark $\tilde\chi_i=\chi_{i2}\sim(3,1,-2/3)$. We discuss its contributions to $R_{D^{(*)}}$ in Sec.~\ref{sec:bctaunu}.

\subsection{Neutral current process: $b\to s\ell\ell$}
\label{sec:bsll}

We briefly outline the most important points from the study in Ref.~\cite{Balaji:2018zna} and refer the interested reader to the publication for further details.
The relevant $SU(4)$ gauge interactions with the fermions
are given by
\begin{align}
	\mathcal{L} = 
	\frac{g_s}{\sqrt{2}}K_{ij} W^\prime_\mu \bar d_i \gamma^\mu P_L \ell_j
	+ \frac{g_s}{\sqrt{2}} K^{*}_{ji} W^{\prime *}_\mu \bar \ell_i \gamma^\mu P_L d_j
\end{align}
where $g_s$
is the $SU(4)_C$ gauge coupling constant and $K$ is the mixing matrix between left-handed charged leptons and down-type quarks as shown in Ref.~\cite{Balaji:2018zna}. As quantum chromodynamics $SU(3)_C$ is embedded in $SU(4)_C$, the coupling $g_s$ is directly defined by the strong gauge coupling.
Here we have defined $\ell$ to include the three charged SM leptons and the three heavy exotic charged lepton mass eigenstates, i.e. $\ell = e,E$.

After integrating out the heavy $W'$ mediator with mass $m_{W^\prime}$ there are new contributions to the Wilson coefficients of $b\to s \ell\ell^\prime$,
	\begin{align}\label{eq: C9 and C10}
	C_9^{sb\ell \ell^\prime} & = - C_{10}^{sb\ell \ell^\prime} =
	\frac{\sqrt2 \pi^2 \alpha_s  }{ V_{ts} V_{tb}^* \alpha_{em}} \frac{K_{s\ell^\prime} K_{b\ell}^*}{G_F m_{W^\prime}^2 } \;.
\end{align}
In the above $\alpha_s = g_s^2(m_{W'})/4\pi$ is the running strong coupling constant and $\alpha_{em} =1/127.9$ denotes the fine-structure constant evaluated at the electroweak scale. $K_{ij}$ are the elements of a CKM-type quark-lepton mixing matrix.
The Wilson coefficients are defined by the effective Lagrangian
\begin{equation}
	\mathcal{L}_{eff} = \frac{4 G_F}{\sqrt{2}} \frac{\alpha_{em}}{4\pi} \sum_{\ell, \ell^\prime} V_{ts} V_{tb}^*  \sum_{i=9,10}
	C_i^{sb \ell \ell^\prime} O^{sb\ell\ell^\prime}_i
	+\mathrm{h.c.}\;,
\end{equation}
where $O_i$ denotes operators with a strange and bottom quark and two charged leptons
\begin{align}
	O_9^{sb\ell\ell^\prime} & = (\bar s\gamma_\mu P_L b ) (\bar \ell \gamma^\mu \ell^\prime) &
    O_{10}^{sb\ell\ell^\prime} & = (\bar s\gamma_\mu P_L b) (\bar \ell \gamma^\mu \gamma_5\ell^\prime) 
\ .
\end{align}

In order to explain the $R_{K^{(*)}}$ anomalies and to avoid stringent constraints from the lepton-flavor-violating  $K_L\to e^\pm\mu^\mp$ decays among others, a particular off-diagonal structure of the CKM-type quark-lepton mixing $K$ matrix is suggested. Considering only the first three columns of the general $K$ matrix, i.e. the part relevant to quark-SM lepton interactions, we adopt the limiting case \footnote{In general $(K_{ij})$ is a $3\times 6$ matrix which satisfies the unitarity condition $K K^\dagger  = 1_{3\times 3}$, where $1_{3\times 3}$ is the $3\times 3$ unit matrix.}
\begin{align}
	K &= \begin{pmatrix}
		0 & 0 & 1 \\
		\cos\theta & \sin\theta & 0 \\
		-\sin\theta & \cos\theta & 0 \\
	\end{pmatrix} \;.
\label{m5}
\end{align}

\subsection{Charged current process: $b\to c\tau \bar \nu$}
\label{sec:bctaunu}
The leptoquark $\tilde\chi$ couples to both charged leptons and
neutrinos
\begin{align}
        \mathcal{L} &=
        - Y_3 \bar {\bf d}_R {\tilde\chi} {S}_L
	- Y_4 \bar {\bf u}_R {\tilde\chi} {\bf e}^c_R
	- \frac{Y_4}{\sqrt{2}} \bar {\bf d}_R {\tilde\chi} ({\bf N}^c_R)
        +h.c.
\end{align}
Starting from this interaction Lagrangian we derive the Wilson coefficients. The neutrino mass eigenstate $n_4$ mixes with the flavor eigenstates
$(N_R^c)_\beta = U_{N_\beta4} n_4 +\dots$ 
and $S_L=U_{S4} n_4+\dots$  where 
$U$ denotes the matrix diagonalizing the neutral fermion mass matrix $U^T M_{\nu,N} U = \mathrm{diag}(m_1,\dots,m_{10})$. The masses of the neutrino mass eigenstates $n_i$ are denoted $m_i$, $i=1,\dots,10$, where $m_{1,2,3}$ denotes the masses of the three active neutrinos, $m_4$ is the mass of the fourth mass eigenstate $n_4$ and $m_{5,\dots,10}$ labels the masses of the mostly heavy sterile neutrinos. We work in the basis where the right-handed charged leptons and the right-handed up-type quarks are given by their mass eigenstates. Then the relevant part of the interaction Lagrangian for $\tilde\chi$ reads 
\begin{align}
        \mathcal{L} &=
        - \left(Y_{d4} \bar{\bf d}^\prime_R n_4 + Y_4 \bar {\bf u}^\prime_R {\bf  e}^{\prime c}_R\right)\tilde\chi 
        +h.c.
\end{align}
with $Y_{d4} = (R_d)^*_{\alpha d} \left[(Y_3)_{\alpha} U_{S4} + \tfrac{(Y_4)_{\alpha \beta}}{\sqrt{2}} U_{N_\beta 4}\right]$, where $R_d$ relates the weak interaction eigenstates ${\bf  d}_R = R_d {\bf d}^\prime_R$ with the mass eigenstate ${\bf d}^\prime_R$. In the following we use $R_d=1$ and drop the primes from the mass eigenstates. 
Integrating out scalar $\tilde\chi$ results in
\begin{align}
        \mathcal{L} 
        & = \frac{Y_{b4} Y_{c\tau}^*}{4 m_{\tilde\chi}^2}  (\mathcal{O}^{cb\tau 4}_{VR} + \mathcal{O}_{AR}^{cb\tau 4})
\end{align}
among other operators. The effective vector $\mathcal{O}_{VR}^{cb\ell\nu}$ and axial-vector $\mathcal{O}_{AR}^{cb\ell\nu}$ operators for a lepton $\ell$ and right-handed neutrino $\nu$ are defined according to Ref.~\cite{Bardhan:2016uhr} as 
\begin{align}
	\mathcal{O}_{VR}^{cb\ell \nu} & = (\bar c\gamma_\mu  b ) (\bar \ell \gamma^\mu P_R \nu) &
	\mathcal{O}_{AR}^{cb\ell \nu} & = (\bar c\gamma_\mu \gamma_5 b ) (\bar \ell \gamma^\mu P_R \nu)
\end{align}
and enter the effective Lagrangian
\begin{equation}
	\mathcal{L}_{eff} = \frac{2 G_F V_{cb}}{\sqrt{2}} \left(
	C_{VR}^{cb\ell\nu } \mathcal{O}^{cb\ell\nu}_{VR}+C_{AR}^{cb\ell\nu } \mathcal{O}^{cb\ell\nu}_{AR}
	\right)\;,
\end{equation}
We may then simply compute the relevant Wilson coefficients required to compute $R_{D^{(*)}}$ which are given by
\begin{equation}
\label{WC}
	C_{VR}^{cb\tau 4 }=C_{AR}^{cb\tau 4}
= \frac{1}{4\sqrt{2} V_{cb} G_F m_{\tilde\chi}^2} \left(y_3 U_{S4} +\frac{Y_{t\mu}}{\sqrt{2}} U_{N_\mu 4}\right) Y_{c\tau}^*\;,
\end{equation}
where we expressed $Y_{b4}$ in terms of the entries of the minimal Yukawa matrix structure defined in Eq.~\eqref{eq:minYuk} and the matrix elements of $U$.

Considering the aforementioned limit where $Y_{ue}w, Y_{c\tau}w, Y_{t\mu}w\gg m_t, m^{*},m_S$, we may compute the mixing angles $U_{S4}$ and $U_{N_\alpha 4}$ for the fourth neutrino state $n_4$ 
\begin{align}
\label{neutrino mixing angles}
	U_{S4} & = \frac{1}{\sqrt{1+2|\tfrac{y_3}{Y_{t\mu}}|^2}} &
	U_{N_e4}&=0 &
    U_{N_\mu4}&=\frac{\sqrt{2}y_3}{Y_{t\mu}}\frac{1}{\sqrt{1+2|\tfrac{y_3}{Y_{t\mu}}|^2}} & 
    U_{N_\tau4}&=0
\end{align}
to leading order.
So we note that with the selected Yukawa structure, the neutrino that participates in the $R_{D^{(*)}}$ anomalies is dominantly a mixture of the singlet $S_L$ and the second state $N_\mu$ in ${\bf N}^c_R$. Substituting the above mixing angles into \eqref{WC} results in 
\begin{equation}
\label{WC simplified}
C_{VR}^{cb\tau4 }=C_{AR}^{cb\tau4}\approx\frac{1}{2\sqrt{2} V_{cb} G_F m_{\tilde \chi}^2}\frac{y_3 Y_{c\tau}^{*} }{\sqrt{1+2 |\tfrac{y_3}{Y_{t\mu}}|^2}}\;.
\end{equation}

As the decay rates are summed over all polarizations and spins, the expressions for the LFU ratios should be invariant after replacing all Wilson coefficients for left-handed currents by right-handed ones and vice versa~\cite{Asadi:2018wea}. Hence, we may use the literature result for left-handed neutrinos \cite{Bardhan:2016uhr} and map them directly to right-handed neutrinos since there is no interference between left- and right-handed operators. The resulting $1\sigma$ ($90\%$C.L.) bounds on $R_{D^{(*)}}$ from Table~\ref{tab:RKRD} can be directly converted to constraints on the right-handed neutrino current Wilson coefficient 
\begin{equation}
    -0.33 (-0.37)\leq C_{VR}^{cb\tau 4 }\leq-0.25 (-0.19)\;.
\end{equation}


\section{Constraints}
\label{sec:constraints}
Several measurements already place constraints on the favored parameter region. In particular $Z$ boson decays to charged leptons, semihadronic B-meson decays, and collider constraints for the leptoquark, cosmological, astrophysical and direct search constraints on the sterile neutrino $n_4$.

\subsection{$Z$ decay constraints}
The new leptoquark $\tilde \chi$ modifies the $Z$ decay width to muons at one-loop level due to the presence of a large Yukawa coupling $Y_{t\mu}$.
Contributions to other leptonic decays of the $Z$ boson are generally small in this model. 
As the leptoquark $\tilde\chi$ only couples to right-handed charged leptons, its contribution can be parametrized by
\begin{equation}
	\mathcal{L} = \frac{g}{\cos\theta_w} \left[\sin^2\theta_w +\delta g_{\mu,R} \right] \bar \mu_R Z_\mu\gamma^\mu \mu_R 
\end{equation}
following Ref.~\cite{Arnan:2019olv},
where $g$ is the $SU(2)_L$ gauge coupling and $\theta_w$ the weak mixing angle. 
For $m_t\ll m_{\tilde\chi}$ the contribution of the leptoquark $\tilde\chi$ can 
simply be written as
\begin{equation}
\label{Z-decay expression}
\delta g_{\mu,R} = 3 \frac{|Y_{t\mu}|^2}{32\pi^2}x_t(1+\log x_t)
    \end{equation}
    to leading order,
where $x_t=m_t^2/m_{\tilde\chi}^2$. The current best experimental bound from precision
electroweak physics comes from the LEP experiments~\cite{ALEPH:2005ab}. We demand that the $Z$-boson coupling to muons is not changed by more than the experimental uncertainty at $(1\sigma)$ [90\% C.L.], i.e.~$\left|\delta
g_{\mu,R}\right|< \delta g_{\mu,R}^{\rm exp} = (1.3)[2.1]\times10^{-3}$, and thus we obtain the constraint
\begin{equation}
\label{Z-decay constraint}
|Y_{t\mu}| \leq \frac{4\pi\sqrt{2 \delta g_{\mu_R}^{\rm exp}}}{\sqrt{3 x_t\left(1+\log x_t\right)}}\;.
    \end{equation}

\subsection{$B\to K\nu\bar\nu$}
Another constraint comes from $B\to K\nu\bar\nu$ which is modified by the leptoquark $\tilde \chi$. It is described by effective operators of the form \cite{Buras:2014fpa} 
\begin{align}
	\mathcal{L} = 2\sqrt{2} G_F V_{tb}V_{ts}^* \frac{\alpha_{em}}{4\pi} \sum_{X=L,R} C_{\nu,X} \bar s \gamma_\mu P_{X} b \bar \nu (1-\gamma_5)\nu\;.
	\end{align}
Integrating out $\tilde \chi$ as before we obtain
\begin{align}
        \mathcal{L} &
	= \frac{(\sqrt{2} y_3U_{S4} + Y_{t\mu}U_{N_\mu 4})^* Y_{c\tau} U_{N_\tau 4}}{4m_{\tilde\chi}^2}  (\bar s \gamma^\mu P_R   b) (\bar n_4 \gamma_\mu P_L n_4)
	\;.
\end{align}
As $|U_{N_\tau 4}|\ll 1$ we find that the new physics contribution $C_{\nu,R}^{\rm NP}$ is very small compared to the SM contribution $C_{\nu,R}^{\rm SM}=-6.38\pm0.06$ and thus $B\to K\nu\bar\nu$ does not provide any competitive constraint.
Similarly, new contributions from the exchange of $\tilde\chi$ to $B\to\pi\nu\bar\nu$ and $K\to \pi\nu\bar\nu$ decay rates are very suppressed due to the assumed Yukawa coupling structure.


\subsection{Collider constraints}
There currently does not exist a plethora of dedicated searches at colliders for the leptoquark $\tilde \chi$ since the chosen Yukawa texture by construction couples only second-generation quarks with third-generation leptons and vice versa. The most common LHC searches are for single generation leptoquarks \cite{Sirunyan:2018vhk, Sirunyan:2018jdk, Sirunyan:2018jdk,Khachatryan:2016jqo, Khachatryan:2014ura, Chatrchyan:2012st, Chatrchyan:2012sv}. The searches are commonly separated into single leptoquark and pair production. The latter is generally independent of the absolute magnitude of the leptoquark Yukawa coupling, because the leptoquarks are produced via strong interactions in a hadron collider, unless the Yukawa couplings are large and substantially contribute to the leptoquark production, while single leptoquark production depends on the Yukawa coupling.  

The model parameter space can be most economically constrained by searches with $\tau c$ or $\mu t$ final states. The $\tilde \chi$ particle can also decay into $b\nu$ due to the coupling $y_3$ being nonzero, but searches for final states with missing transverse energy are typically less sensitive. For the chosen mass range of the scalar leptoquark $\tilde \chi$ imposing the constraints from searches with third-generation scalar leptoquarks decaying into a tau lepton and a $b$ quark such as the analysis in Ref.~\cite{Sirunyan:2018jdk} does not pose any additional constraint on the parameter space over the $Z$ decay constraint, although the sensitivity on the quark is markedly improved due to $b$ tagging of the final state jets. However, a mixed 1-3 generation leptoquark search with final states $\mu t$ analogous to the third-generation search in Ref.~\cite{Khachatryan:2015bsa} could strengthen the limits on the $Y_{t \mu}$ coupling significantly in the future. Of course more complicated Yukawa textures for $Y_3$ and $Y_4$ (particularly the diagonal entries), can be chosen and constrained with the aforementioned single generations searches, however we do not consider these more complicated parametrizations in this work for the sake of brevity.

Finally there are constraints from the single $\tau$-lepton + MET searches. The authors of Ref.~\cite{Greljo:2018tzh} reinterpreted the searches for a heavy charged gauge boson from sequential SM (SSM) resonance searches of the ATLAS~\cite{Aaboud:2018vgh} and CMS~\cite{Sirunyan:2018lbg} experiments as constraints on models explaining $R_{D^{(*)}}$. In particular, the leptoquark $\tilde\chi$ with purely right-handed couplings has been studied and the study finds that leptoquark masses above 2 TeV are excluded at more than $2\sigma$. At face value this constrains the leptoquark $m_{\tilde \chi}$ to be lighter than 2 TeV. As the study was based on an older best fit to the $R_{D^{(*)}}$ anomalies further away from the SM, the current constraint for heavy charged gauge bosons from SSM resonance searches is relaxed and heavier masses are allowed. However, the precise value of the current constraint requires a new study. In the following results discussion, we thus consider leptoquark masses up to 3 TeV and caution the reader that the SSM resonance search poses a constraint on the heaviest allowable leptoquark masses according to the study shown in Ref.~\cite{Greljo:2018tzh}.

\subsection{Constraints on the sterile neutrino}
The sterile neutrino $n_4$ as defined would be produced in the early Universe. The dominant decay modes are $n_4 \to \nu_\alpha f \bar f$ with $f=\nu_\beta, e^-, \mu^-$ for masses $m_4\leq 1$ GeV. These decays are mediated by the $Z$ boson and thus the decay rate depends quadratically on the $\nu_\alpha-n_4$ mixing matrix element $|U_{\alpha 4}|^2$. In the limit of vanishing final state lepton masses, the decay rate of $n_4\to\nu_\alpha \bar f f$ is given by ~\cite{Lee:1977tib,Pal:1981rm,Barger:1995ty}
\begin{equation}
\Gamma(n_4\to \nu_\alpha \bar ff ) = \frac{G_F^2 m_4^5}{96\pi^3} |U_{\alpha 4}|^2\;.
\end{equation}
For $2m_\mu \geq m_4\gg 2m_e$ the lifetime is given 
\begin{equation}
\tau = \Gamma(n_4\to \nu\bar f f)^{-1} = \frac{96\pi^3} {4G_F^2 m_4^5 \sum_\alpha|U_{\alpha 4}|^2}
\simeq 0.04 s \left(\frac{100 \mathrm{MeV}}{m_4}\right)^5 \left(\frac{10^{-5}}{\sum_\alpha |U_{\alpha 4}|^2}\right)
\;.
\end{equation}
Big bang nucleosynthesis (BBN) poses a constraint on the lifetime of $n_4$, since the abundances of the light elements agree well with the standard cosmological model. Thus in order to avoid any changes to the standard BBN, the sterile neutrino $n_4$ has to decay and its decay products thermalize, before BBN. If the lifetime of $n_4$ is shorter than $\tau< 0.1s$, this condition can be satisfied as it has been shown in Refs.~\cite{Dolgov:2000pj,Dolgov:2000jw}. This translates into a bound
\begin{equation}
\label{BBN constraint}
m_4  \gtrsim 87\, \mathrm{MeV} \left(\frac{10^{-5}}{\sum_\alpha |U_{\alpha 4}|^2}\right)^{1/5}
\;.
\end{equation}

Similarly, sterile neutrinos can be produced in supernovae. The arguments of Ref.~\cite{Dolgov:2000pj} imply that the duration of the SN 1987A neutrino burst excludes mixing angles $3\times 10^{-8} < \sin^22\theta < 0.1$ for sterile neutrinos $m_4\lesssim 100$ MeV.

Finally, sterile neutrinos can be searched for at terrestrial experiments. In particular the fixed-target experiments NOMAD~\cite{Astier:2001ck} and CHARM~\cite{Orloff:2002de} placed limits on the mixing angle of sterile neutrinos with $\tau$ neutrinos, which further constrains the allowed parameter space, as discussed in Sec.~\ref{sec:results}. 

Together this puts a lower bound on the sterile neutrino mass of $m_4\geq100$MeV.

\section{Results}
\label{sec:results}

As the explanations for the $R_{K^{(*)}}$ and $R_{D^{(*)}}$ anomalies are mostly independent, we first discuss the explanation of $R_{K^{(*)}}$, which sets the scale of the $SU(4)_C\times SU(2)_L\times SU(2)_R$ symmetry breaking. As a second step, we present the favored region for the anomalies $R_{D^{(*)}}$, before finally discussing its predictions for the sterile neutrino $n_4$ in the process $b\to c\tau n_4$. 

\subsection{$R_{K^{(*)}}$}
\begin{figure}[tb!]
    \centering
    \includegraphics[width=0.7\linewidth] {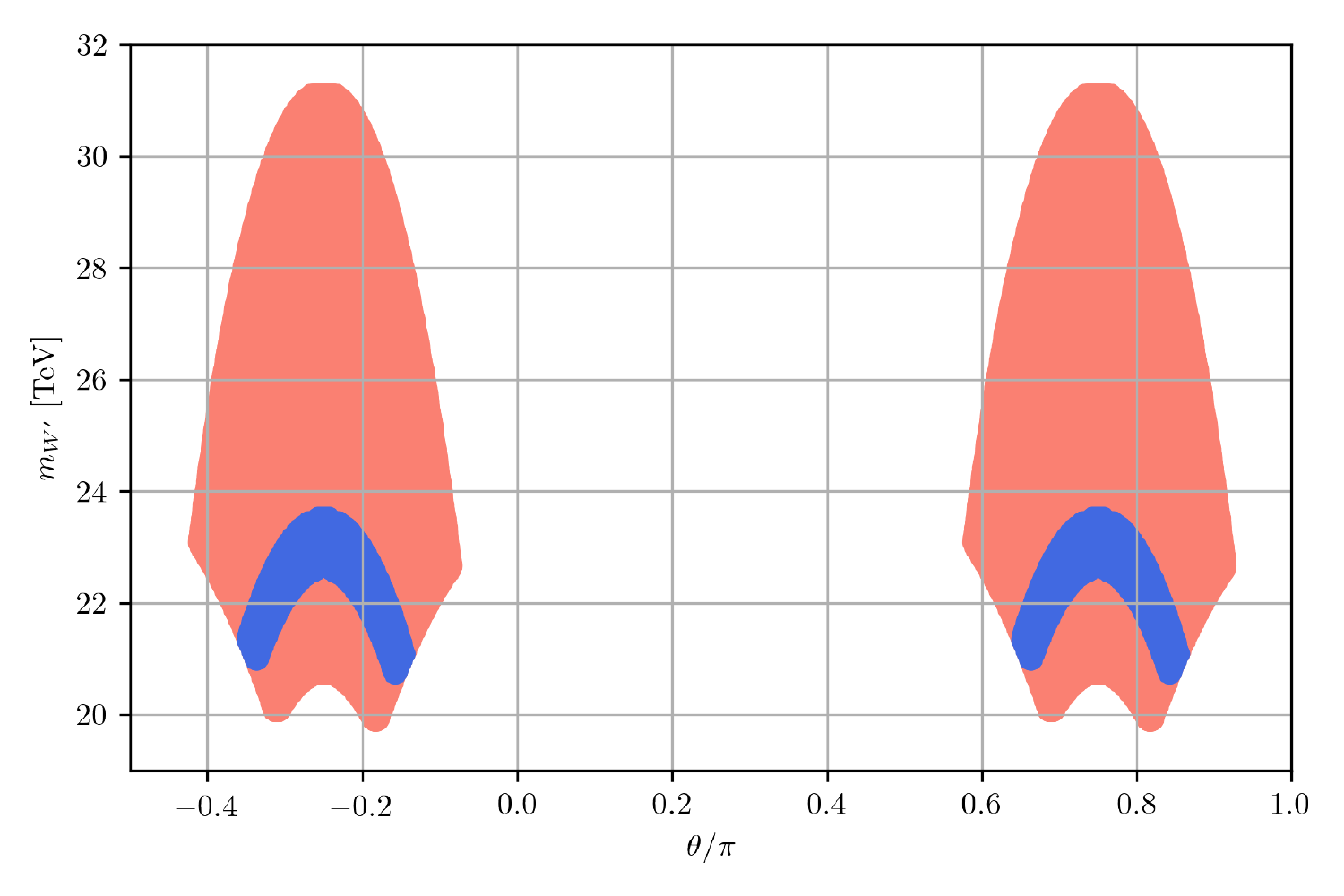}
    \caption{The favored parameter regions compatible with the current experimental limits from 
${B^+}\to {K^+}\mu^- e^+$, ${B^+}\to {K^+} e^- \mu^+$. 
Shown are the 1$\sigma$ (blue) and 90\% confidence level (red) bands suggested by the measured $R_{K}$ and $R_{K^*}$ ratios. 
}
    \label{fig:Allowable regions}
\end{figure}

We follow our previous analysis~\cite{Balaji:2018zna} and identify the favored region of parameter space for the model using the \texttt{flavio} package \cite{david_straub_2018_1326349} and tree-level analytical estimations where appropriate. 
The $1 \sigma$ (90\% C.L.) favored parameter region is defined by
the values of the vector leptoquark mass $m_{W'}$ and the quark-SM lepton mixing angle $\theta$ [see Eq.~\eqref{m5} for its definition] which satisfy $R_K = 0.846^{+0.062}_{-0.056}$
($R_K = 0.846^{+0.102}_{-0.091}$), $R_{K^*} = 0.685^{+0.122}_{-0.083}$ ($R_{K^*} = 0.685^{+0.201}_{-0.137}$) and also satisfy the current  90\% C.L. experimental
limits $BR({B^+}\to {K^+}\mu^- e^+)<7\times10^{-9}$ and $BR({B^+}\to {K^+} e^-
\mu^+)<6.4\times10^{-9}$ \cite{Tanabashi:2018oca}. Other processes currently do not constrain the parameter region as we discussed in Ref.~\cite{Balaji:2018zna}.

Figure~\ref{fig:Allowable regions} shows the favored region of parameter space in the $m_{W'}$ leptoquark mass versus the $\theta$ mixing angle plane. Compared to Ref.~\cite{Balaji:2018zna} the mass of $W'$ is larger, because the experimentally observed value of $R_K$ has since  moved closer to the SM prediction with smaller error bars and therefore the $1\sigma$ region is smaller. 
The favored range of $\theta$ is approximately between $[-\frac{\pi}{2},0]$ or $[\frac{\pi}{2},\pi]$ and $m_{W'}$ between $[20, 31]$ TeV. The identical nature of the two adjacent regions can be understood from the invariance of the relevant branching ratios under the transformation $\theta \to\theta+\pi$.
The constraints from $B\to K e^\pm \mu^\mp$ lead to the wedge-shape form at the bottom of each favored region. 

\begin{figure}[tb!]
\centering
        \includegraphics[width=0.48\linewidth]{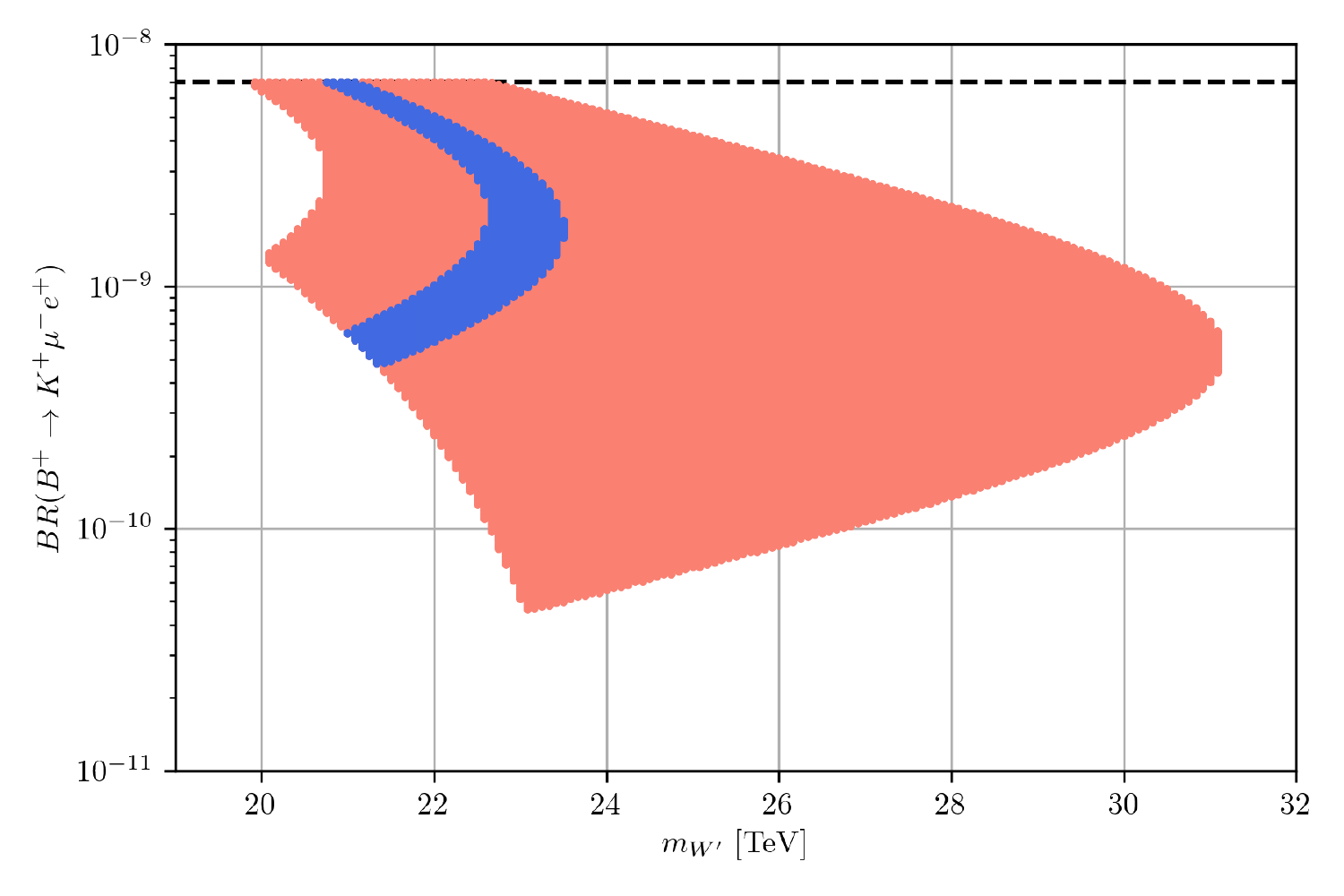}
        \includegraphics[width=0.48\linewidth]{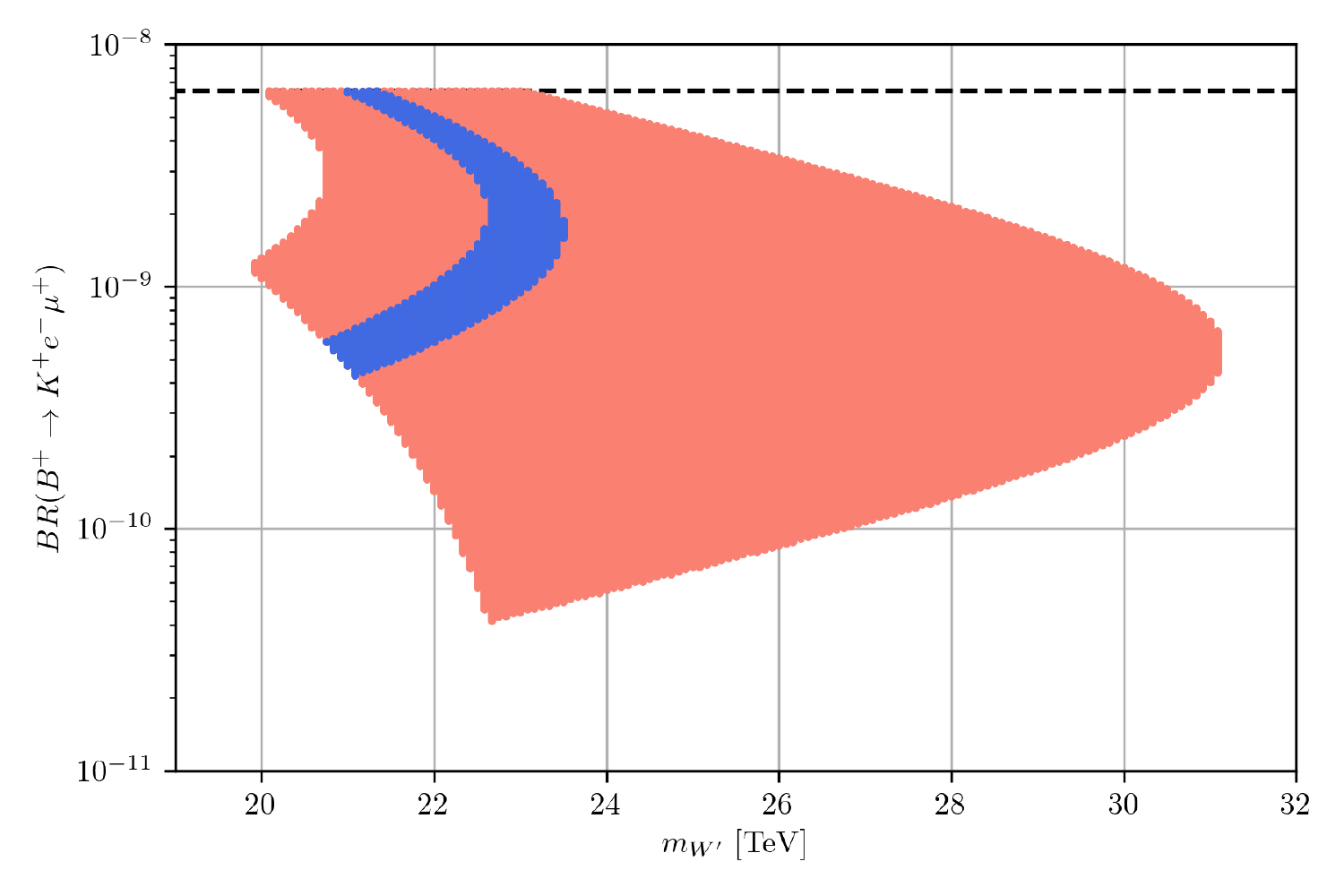}
	\caption{Expectation for $BR({B^+}\to {K^+}\mu^- e^+)$ (left) and $BR({B^+}\to {K^+} e^- \mu^+)$ (right)
 for the favored parameter region identified in Figure \ref{fig:Allowable regions}. 
The black dashed lines correspond to the current experimental 90\% confidence-level upper bounds on these branching fractions. 
}
    \label{fig:BR Predictions}
\end{figure}

Figure \ref{fig:BR Predictions} shows the predicted range for the branching ratios of the lepton-flavor-violating rare decays ${B^+}\to {K^+}\mu^- e^+$ and ${B^+}\to {K^+} e^- \mu^+$ processes. The two processes probe different ranges of $\theta$ values and are thus complementary: While $B^+ \to K^+\mu^- e^+$ is sensitive to $\sin^2\theta \approx 1$, $B^+ \to K^+\mu^+ e^-$ is sensitive to $\cos^2\theta
\approx 1$. LHCb is expected to further improve its sensitivity and to probe the two branching ratios of $B^+\to K^+e^\pm \mu^\mp$ at the level of $10^{-9}$~\cite{Bediaga:2018lhg}.

In addition to further improvements to $B^+ \to K^+ \mu^{\pm} e^{\mp}$ this
leptoquark contributes to lepton-flavor-violating rare tau lepton decays such as $\tau\to
{K_s}\ell$, $\ell = e, \mu$ and the leptonic $B_s$ decays $B_s\to \ell^-\ell^{\prime+}$, $\ell,\ell^\prime=e,\mu$ as shown in Ref.~\cite{Balaji:2018zna}. However, the additional contributions are below the current experimental sensitivity and thus we do not show
these predictions for the sake of brevity and refer the interested reader to Ref.~\cite{Balaji:2018zna} for further details.


\subsection{$R_{D^{(*)}}$}
Using Eq.~\eqref{WC simplified} along with the $1\sigma$ and $90\%$ C.L. $R_{D^{(*)}}$ constraints on the Wilson coefficient $C_{VR}^{cb\tau 4}$ we may derive the allowable parameter region for the model which satisfies the anomalies. We restrict ourselves to placing bounds on the $1\sigma$ and $90\%$ C.L. $R_{D^{(*)}}$ region. Choosing the minimal Yukawa texture described in Sec.~\ref{sec:YukawaSector} for $Y_3$ and $Y_4$  constrains the parameter region in $Y_{c\tau}$, $Y_{t\mu}$, $y_3$ and $m_{\tilde\chi}$ space.

We limit $Y_{ue}\simeq0.1$ for our parameter scans to ensure that the lightest flavor exotic charged lepton $E$ mass is larger than $\simeq 1$ TeV for scales larger than $w\simeq 10$ TeV, and this coupling does not affect the neutrino states $S$ and $N_\mu$ that participate in the anomaly and is therefore not important in constraining the model's allowable region. The Yukawa couplings of interest must also satisfy perturbativity requirements such that $0 \leq Y_{c\tau}\leq 4\pi$, $0 \leq Y_{t\mu}\leq 4\pi$ while $-4\pi \leq y_3 \leq 0$ in order to obtain Wilson coefficient $C_{VR}^{cb\tau 4}$ with the correct sign. The VEV $w=26.7$ TeV was chosen for our parameter scans as this is a favored central value for the $m_{W'}\simeq23$ TeV gauge boson mass scale which explains the $R_{K^{(*)}}$ anomalies. This fixes the lightest exotic vectorlike lepton mass to $Y_{ue}w\simeq 2.7$ TeV which easily evades the LEP constraints for heavy charged leptons \cite{Tanabashi:2018oca}.

We also set $m_S=2m_{\mu}$ as this acts as an upper bound on the fourth neutrino mass participating in $b\to c\tau \nu$. This value is chosen because it ensures that the sterile neutrino $n_4$ decays before BBN. The analytical approximation for the mixing angles $U_{S4}$ and $U_{N_\mu 4}$ in Eq.~\eqref{neutrino mixing angles} and subsequently Eq.~\eqref{WC simplified} is respected as well as ensuring that the new neutrino mass is light enough that it does not introduce too much phase space suppression in the decay $b\to c\tau n_4$. Consequently in our parameter scan we find the fourth neutrino mass to be lighter than the $2m_\mu$, but heavier than 100 MeV after imposing all constraints, which ensures that it is still significantly heavier than the active neutrinos but sufficiently lighter than the $B$ meson. The parameter ranges of the relevant new physics parameters detailed above are summarized in Table~\ref{tab:parameter scan} for convenience.

\begin{table}[htb!]
\centering
	\begin{minipage}{8cm}
	\begin{ruledtabular}
		\begin{tabular}{cc}
		Parameter &  Value \\ \hline
		$u_{1}$& 0\\
		$v_{12}$& 0\\
		$u_2^2+v_{21}^2$& $1/(2\sqrt{2}G_F)\simeq (174 \mathrm{GeV})^2$\\
		$w$ & $26.7$ TeV  \\
        $m_{\tilde \chi}$ & $[0.8,3]$ TeV  \\
		$m_S$ & $2 m_\mu$  \\
        $y_{3}$ & $[-4\pi,0]$ \\	
        $Y_{ue}, Y_{c\tau}, Y_{t\mu}$ & 0.1, $[0,4\pi]$, $[0,4\pi]$\\
\end{tabular}
\end{ruledtabular}
\end{minipage}
\caption{Parameter ranges for the new physics model parameters used in the numerical scans.}
\label{tab:parameter scan}
\end{table}

\begin{figure}[phbt!]
    \centering
        \includegraphics[width=0.48\linewidth]{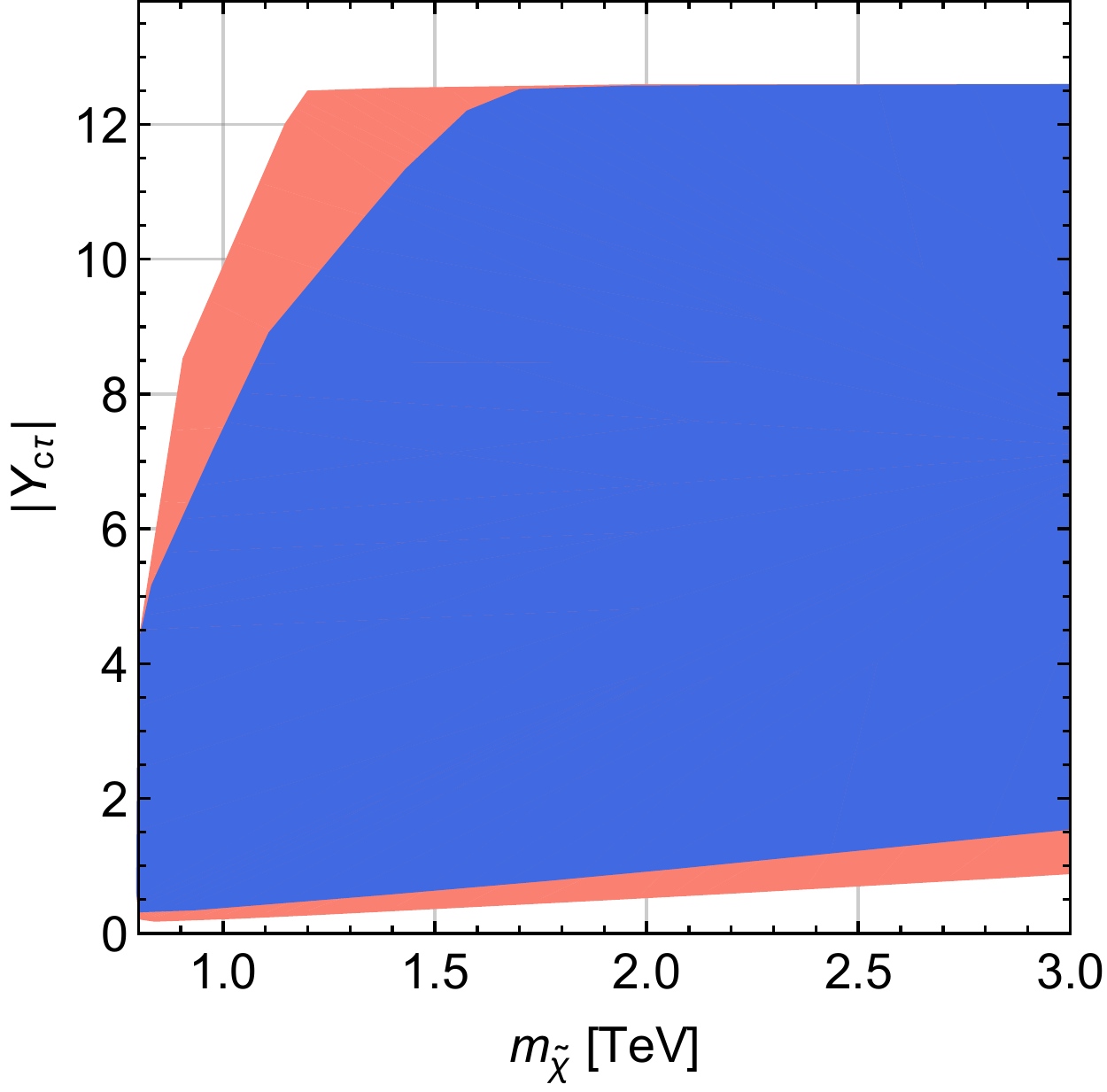}
        \includegraphics[width=0.48\linewidth]{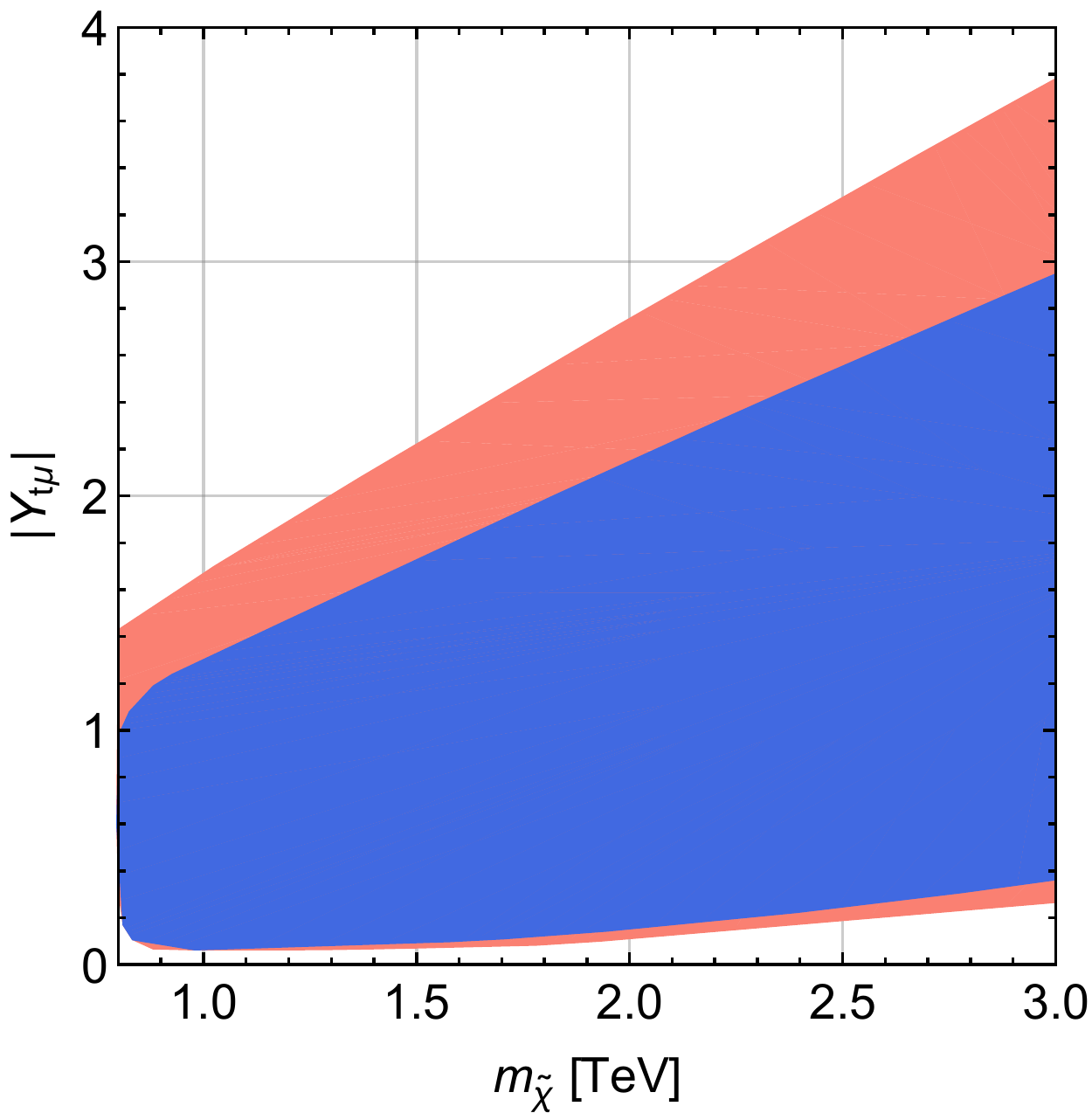}

        \includegraphics[width=0.48\linewidth]{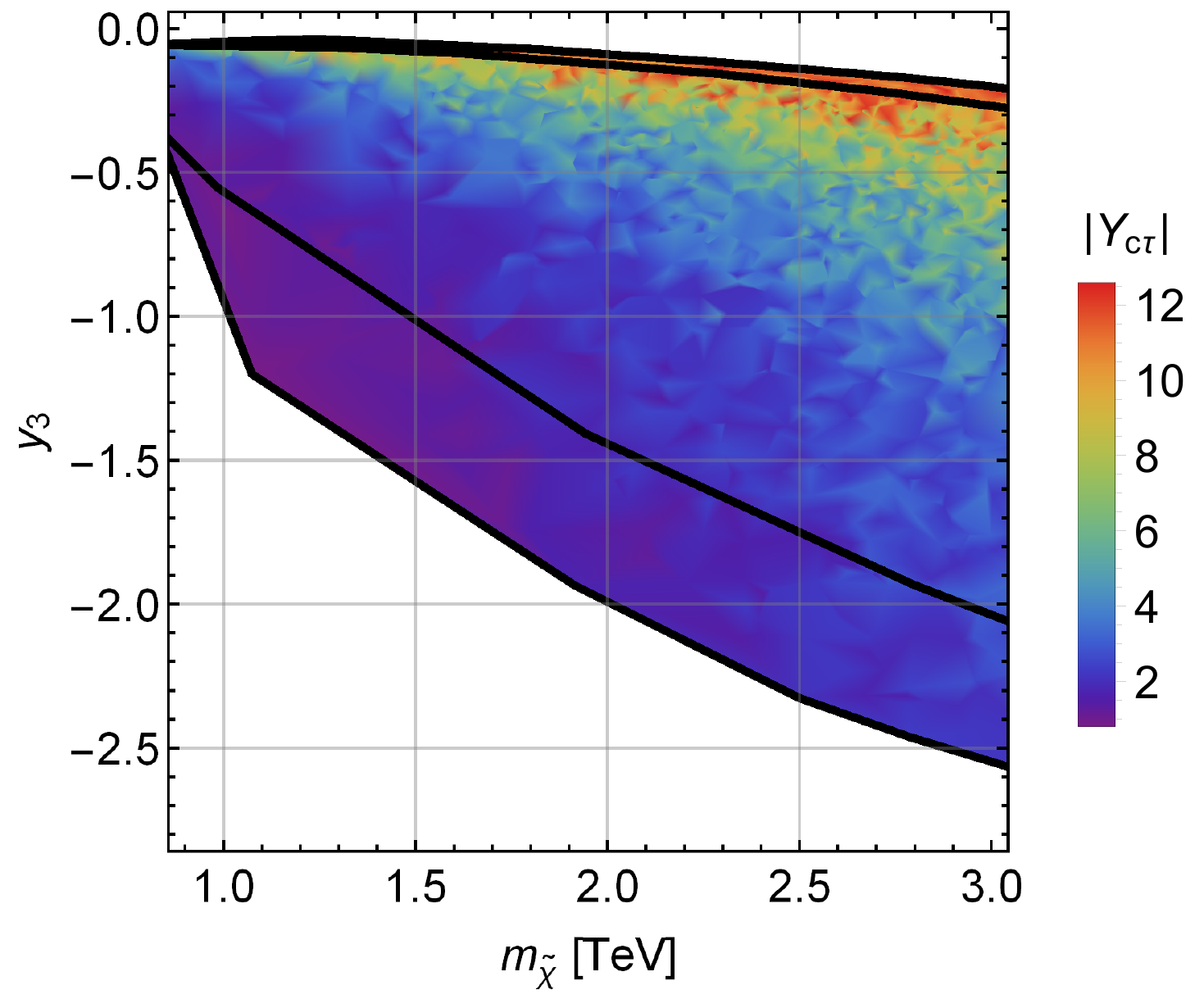}
        \includegraphics[width=0.48\linewidth]{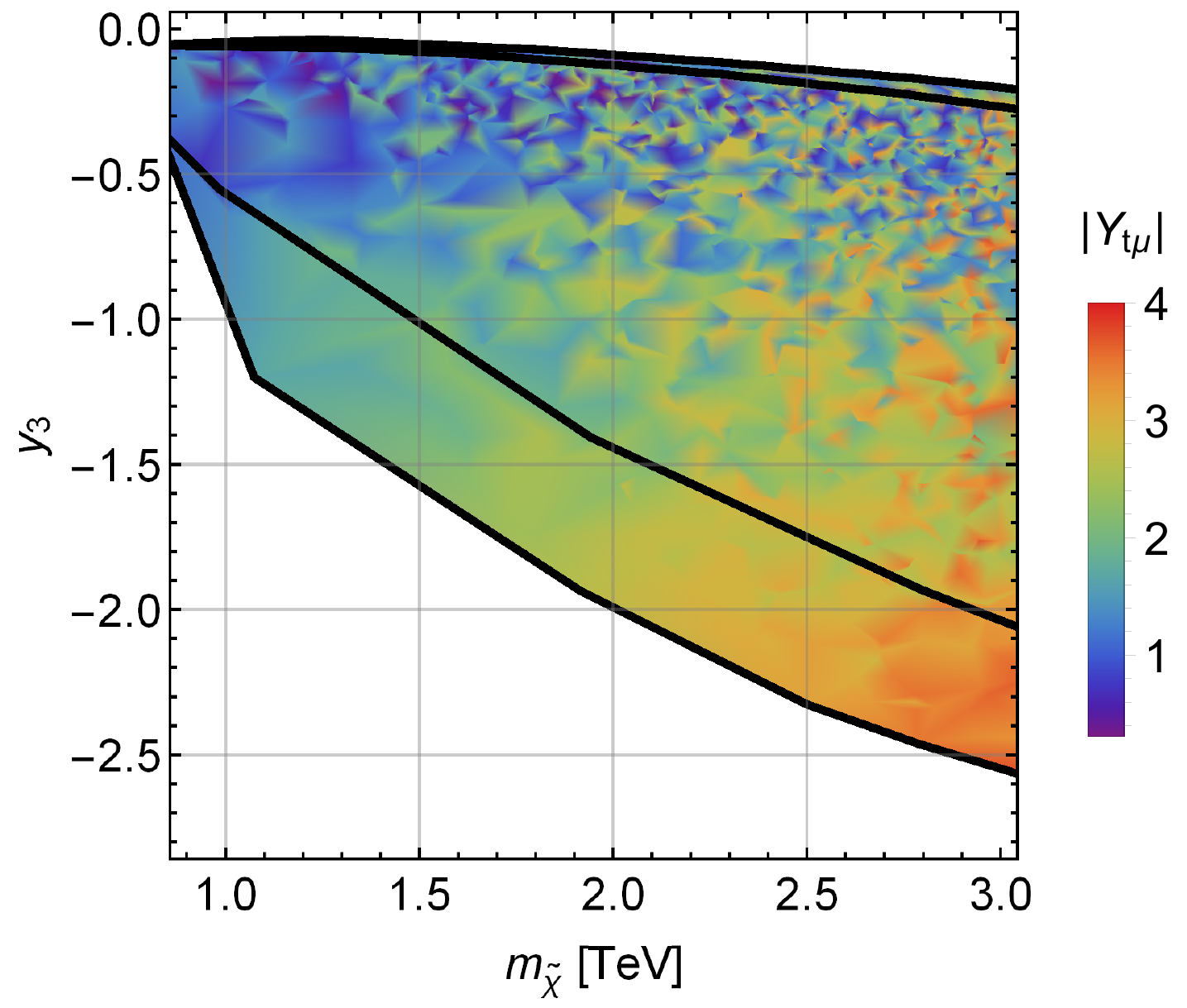}
\caption{\small The top two panels show the allowable $R_{D^{(*)}}$ 1$\sigma$ (blue) and 90\% confidence level (red) parameter regions for the Yukawa couplings $Y_{c\tau}$ (top left) and $Y_{t\mu}$ (top right). The parameter region is displayed over the $0.8 \leq m_{\tilde\chi} \leq 3$ TeV range, which is of immediate interest in current and future TeV scale collider searches. The Yukawa couplings are also restricted to be $\leq 4\pi$ to remain in the perturbative regime. The $Z\to \mu\mu$ $1\sigma$ and 90\% confidence new physics coupling correction $\delta g _{\mu R}$ constraints also enforce an important upper cutoff on $Y_{t\mu}$ as seen in the right-hand side allowable regions. In the lower panels we show density plots which are a result of our numerical scan with the additional BBN constraint shown in Eq.~\eqref{BBN constraint} and the SN 1987A~\cite{Dolgov:2000pj,Dolgov:2000jw} and CHARM~\cite{Orloff:2002de} neutrino mixing constraints, where $y_{3}$ is a function of $m_{\tilde{\chi}}$, $Y_{c\tau}$ (bottom left) and $Y_{t\mu}$ (bottom right). The region contained within the inner and outer black boundaries corresponds to the $1\sigma$ and 90\% confidence level regions respectively. Note that the sharp edges and color discontinuities are due to limitations in numerical sampling and not physical effects.}
\label{f2} 
\end{figure}

We also ensure that the leptonic mixing parameters and the neutrino mass squared differences satisfy the $3\sigma$ ranges from the latest global fit by the NuFIT collaboration~\cite{Esteban:2018azc}: $0.275\leq \sin^2\theta_{12}\leq 0.350$, $0.427 \leq \sin^2\theta_{23} \leq 0.609$, $0.02046\leq\sin^2\theta_{13}\leq0.02440$ and the solar and atmospheric mass squared differences $6.79\leq \frac{\Delta m_{21}^2}{10^{-5}eV^2}\leq 8.01$, $2.432\leq \frac{\Delta m_{3\ell}^2}{10^{-3}eV^2}\leq 2.618$. 
We also impose the $3\sigma$ unitarity deviation bound derived from $|2\eta_{\alpha \beta}|$ as shown in Ref.~\cite{Agostinho:2017wfs} on $|UU^\dagger|$. The allowed regions are then constrained by the combination of Yukawa coupling ranges in conjunction with the $Z\to\mu\mu$ constraint in Eq.~\eqref{Z-decay constraint} and the $C_{VR}^{cb\tau 4}$ constraint in Eq.~\eqref{WC simplified}, which can be easily plotted analytically along with the perturbative boundaries.

Figure~\ref{f2} shows the viable parameter ranges for the Yukawa couplings $Y_{c\tau}$, $Y_{t\mu}$, and $y_3$ as a function of the leptoquark mass $m_{\tilde \chi}$. We find that for small $Y_{t\mu}$ we require large $Y_{c\tau}$ and vice versa which is what we expect from inspecting Eq.~\eqref{WC simplified}. It should be noted that more complicated Yukawa textures for $Y_4$ and $Y_3$ are indeed permissible as mentioned earlier. But our selection is motivated by maintaining simplicity and reducing the number of free parameters in the theory. If the $R_{D^{(*)}}$ anomalies persist and new stronger constraints become available reducing the parameter space of this chosen texture, other more elaborate ones can indeed be explored.

\subsection{Prediction for neutrino mixing and mass of $n_4$ }

\begin{figure}[bt!]
\vspace{-0.5cm}
\centering
\includegraphics[width=0.7\linewidth]{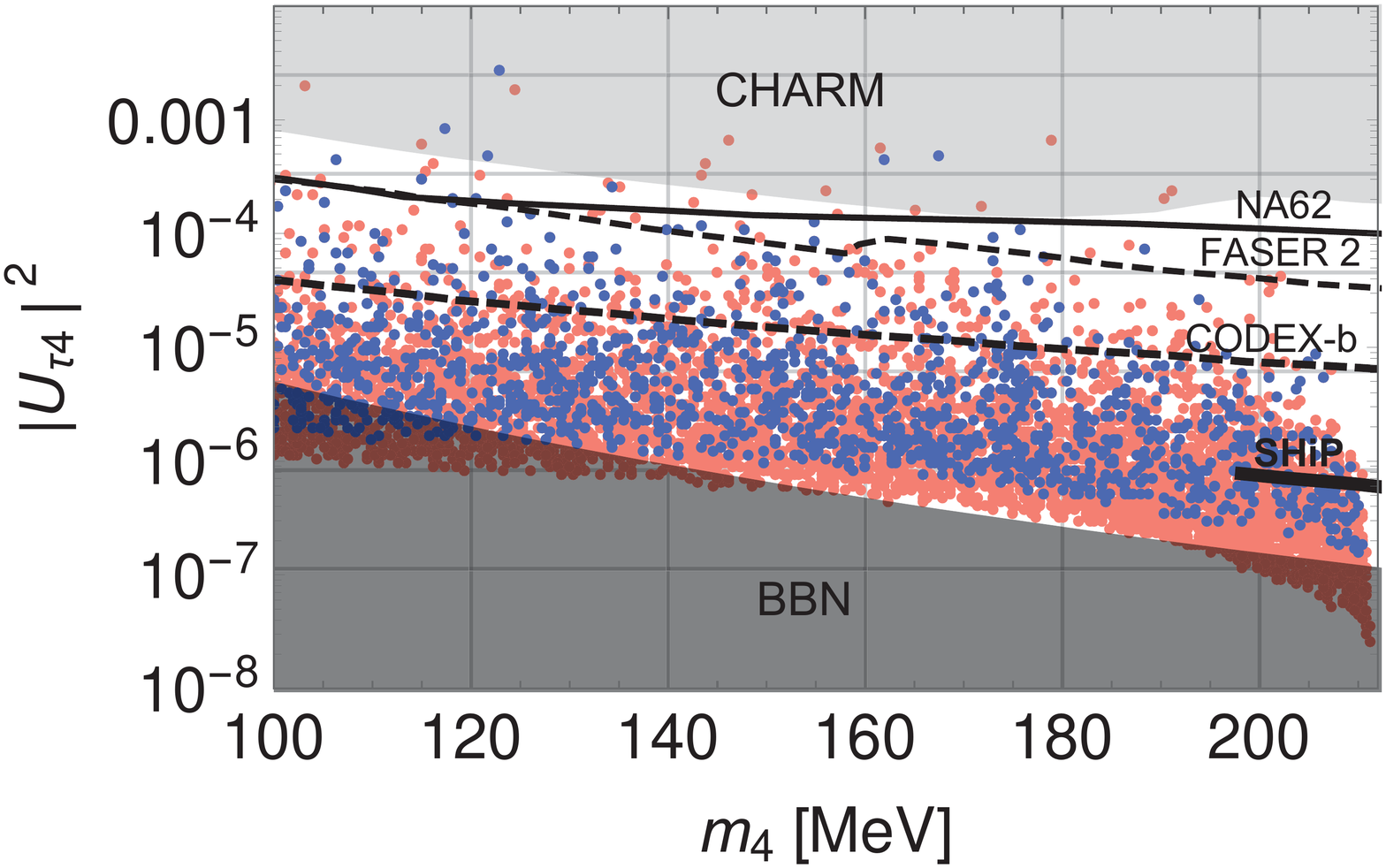}

\caption{\small Prediction for the mixing between the fourth neutrino mass eigenstate $n_4$ participating in the $R_{D^{(*)}}$ anomalies with the dominant active neutrino flavor $\tau$ as a function of its mass $m_{4}$. The blue and red regions correspond to the $1\sigma$ and 90\% confidence level regions respectively while the bottom black shaded region corresponds to the BBN exclusion bound shown in Eq.~\eqref{BBN constraint} and the top bound shown in gray comes from the CHARM experiment. The lines show projected upper bounds for the NA62 (black), FASER 2 (dashed black), CODEX-b (thick dashed black) and SHiP (thick black) experiments from top to bottom respectively.
}

\label{f3} 
\end{figure}

We may additionally predict the mixing of the fourth neutrino mass eigenstate $n_4$ with the active neutrinos. In our numerical scan we find that the mixing matrix elements $U_{e4}$ and $U_{\mu4}$ are negligibly small, $|U_{e4}|^2\lesssim 10^{-11}$ and $|U_{\mu 4}|^2\lesssim 10^{-10}$, due to $y_3$ being the only nonzero element in the chosen texture for $Y_3$. In Figure~\ref{f3} we show the allowable region of parameter space as a function of $|U_{\tau4}|^2$ vs the sterile neutrino mass $m_4$. The BBN constraint from Eq.~\eqref{BBN constraint} results in a lower bound on the mixing matrix element $|U_{\tau 4}|^2$ as a function of the sterile neutrino mass. The duration of the neutrino burst of SN 1987A imposes a lower bound on the sterile neutrino mass $m_4\geq100$ MeV and thus we only show sterile neutrino masses heavier than $100$ MeV. \\\\
In this study, we focus on light sterile neutrino masses satisfying $m_4\leq 2m_\mu$, because the contribution to $R_{D^{(*)}}$ is phase space suppressed for a heavy sterile neutrino $n_4$. Indeed larger neutrino masses could still be kinematically accessible and interesting to study in the light of the MiniBooNE excess as proposed in Ref.~\cite{Fischer:2019fbw}. However we do not analyze such cases in this work. There are additional constraints coming from the NOMAD~\cite{Astier:2001ck} and CHARM~\cite{Orloff:2002de} fixed-target experiments, the stronger of which comes from the CHARM experiment which we also show in Fig.~\ref{f3}. It is also of interest to compare the projected experimental sensitivities for $n_4$, i.e. a sterile neutrino which almost exclusively mixes with $\nu_\tau$, with proposals of future experiments including NA62~\cite{Dobrich:2018ezn}, FASER~\cite{Feng:2017uoz}, CODEX-b~\cite{Gligorov:2017nwh} and SHiP~\cite{Bonivento:2013jag}. The contours have been extracted from Ref.~\cite{Ariga:2018uku}. We note that the SHiP contour only starts at around $m_4\simeq191$ MeV coinciding with the mass splitting between the $D_s^\pm$ meson mother and tau lepton daughter.

\section{Conclusion}
\label{sec:conclusions}
We have proposed a chiral Pati-Salam theory with gauge group $SU(4)_C\times
SU(2)_L\times SU(2)_R$ which is capable of explaining the
$R_{D^{(*)}}$ anomalies with new scalar leptoquarks and the $R_{K^{*}}$ anomalies via $SU(4)$ gauge boson leptoquarks. The model is consistent with experimental constraints, including the fermion mass spectrum, modifications to leptonic
$Z$-boson decays via the new scalar leptoquark, $B\to K\nu \bar{\nu}$ as well as the
best available LHC constraints for single and pair production searches of
leptoquarks at the LHC and other new particles. New physics coming from the gauge sector via a spectrum
of colored leptoquarks with charge $\frac{2}{3}e$ also satisfies the best
available constraints from lepton number violating searches such as $B^{+} \to
K^{+}\mu^{\mp}e^{\pm}$. These gauge bosons couple
in a chiral manner to the familiar quarks and leptons and interfere
with standard model weak processes. 

Both the scalar and massive vector leptoquarks originate from one scalar
multiplet $\chi$ which breaks the Pati-Salam group to the SM group, $SU(4)_C \times SU(2)_R\to SU(3)_C\times U(1)_Y$, at a scale of $\langle\chi_{41}\rangle\equiv w\gtrsim 20$ TeV. 
As already discussed in Ref.~\cite{Balaji:2018zna} the explanation of the $b\to s\ell\ell$ anomalies originates from an equal and opposite tree-level correction to muons and electrons and thus can be tested at the LHCb and Belle II experiments by measuring both lepton flavor-conserving and lepton flavor-violating processes $b\to s \ell \ell^\prime$ and similarly $B_s\to \ell\ell^\prime$, when increased statistics become available. The $R_{D^{(*)}}$ anomalies can be explained using a simple Yukawa texture with only three free parameters, although more complex Yukawa structures are also feasible. There is an intricate relation between the lepton mass spectrum, particularly neutrino mass spectrum, and the $R_{D^{(*)}}$ anomalies. One of the striking signatures is a light sterile neutrino with dominant mixing with tau neutrinos. We constrain the model parameter space using the strong bounds on active-sterile neutrino mixing from big bang nucleosynthesis in conjunction with the supernova SN 1987A and CHARM experiments. Additionally, we make predictions for the sterile neutrino properties which can be probed in future searches such as the proposed NA62, FASER, CODEX-b and SHiP experiments.

\begin{acknowledgments}

We thank Robert Foot for useful discussions and contributions in the early stages of this work. This work has been supported in part by the Australian Research Council (ARC).
\end{acknowledgments}
\appendix


\section{Embedding of SM particle representations}
In the following we list how the different SM and exotic fields are embedded in the $SU(4)_C\times SU(2)_L\times SU(2)_R$ multiplets.
The fermions are decomposed as follows in terms of SM fields:
\begin{align}
	q_{Li\alpha}\equiv Q_{Li\alpha}&\sim (3,2,\frac13)&
	L_{L\alpha}\equiv Q_{L4\alpha}&\sim (1,2,-1)\\
	u_{Ri}\equiv Q_{Ri1}&\sim (3,1,\frac23)&
	d_{Ri}\equiv Q_{Ri2}&\sim (3,1,-\frac13)
			    \\\nonumber
			    \nu_{R}\equiv Q_{R41}&\sim (1,1,0) &
	E_{R}\equiv Q_{R42}&\sim (1,1,-2)\\
	E_L^c\equiv f_{R11} & \sim(1,1,2) & 
	\frac{N_R}{\sqrt{2}}\equiv f_{R(12)} & \sim(1,1,0) &
	e_R\equiv f_{R22} & \sim(1,1,-2) 
	\;.
\end{align}
Note that due to the convention $T_{(ab)}=\frac12(T_{ab}+T_{ba})$ there is a factor $\sqrt{2}$ in the definition of $N_R=\sqrt{2} f_R$ to obtain the correct field normalization of $N_R$.
The scalars can be decomposed in terms of SM fields as follows:
\begin{align}
	\chi_{41}&\sim(1,1,0) &
	\chi_{42}&\sim(1,1,-2) \\\nonumber 
	\chi_{i1} & \sim (3,1,\frac43) &
	S_1^*=\tilde \chi\equiv\chi_{i2} & \sim (3,1,-\frac23) 
\\
	\tilde H_{1\alpha}\equiv \phi_{\alpha 1} &\sim (1,2,1) &
	H_{2\alpha}\equiv \phi_{\alpha 2} &\sim (1,2,-1)
	\\
	R_2 = \Delta_{i\alpha 11} & \sim (3,2,\frac73) &
	\tilde R_2 = \Delta_{i\alpha (12)} & \sim (3,2,\frac13) &
	 \Delta_{i\alpha 22} & \sim (3,2,-\frac53) \\
	 H_{3\alpha}=\Delta_{4\alpha 11} & \sim (1,2,1) &
	 H_{4\alpha}=\Delta_{4\alpha (12)} & \sim (1,2,-1) &
	 \Delta_{4\alpha 22} & \sim (1,2,-3) 
	 \;.
 \end{align}
Thus there are in total four electroweak doublet scalars with hypercharge $\pm1$, one electroweak doublet scalar with hypercharge $-3$, three leptoquarks and another exotic colored scalar which does not couple to quarks and leptons at the renormalizable level.


\bibliography{refs.bib}
\end{document}